# Reputation Algebra for Cloud-based Anonymous Data Storage Systems

Harsh N. Thakker, Mayank Saha and Manik Lal Das

Dhirubhai Ambani Institute of ICT
Gandhinagar – 382007, Gujarat, India
{harshnthakker, saha.mayank}@gmail.com
maniklal_das@daiict.ac.in

**Abstract.** Given a cloud-based anonymous data storage system, there are two ways for managing the nodes involved in file transfers. One of them is using reputations and the other uses a micropayment system. In reputation-based approach, each node has a reputation associated with it, which is used as a currency or feedback collection for file exchange operations. There have been several attempts over the years to develop a strong and efficient reputation system that provides credibility, fairness, and accountability. One such attempt was the Free Haven Project that provides a strong foundation for cloud-based anonymous data storage systems. The work proposed in this paper is motivated by the Free Haven Project aimed at developing a reputation system that facilitates dynamic operations such as adding servers, removing servers and changing role of authorities. The proposed system also provides algorithm for scoring and maintaining reputations of the servers in order to achieve credibility, accountability and fairness.

## 1 Introduction

Cloud-based data storage systems have gained popularity in recent years. Cloud storage involves storing large volumes of data on multiple servers (virtual or physical), which has several advantages over traditional data storage [1], [2]. Although cloud storage provides greater flexibility of accessing data on demand basis from anywhere – anytime, cloud storage services facilitate more importance on persistence of storage rather than accessibility. Based on nature of applications, cloud data access can provide an interesting feature like anonymous data storage and access as well. An anonymous *data storage* system was proposed in [3], known as *The Free Haven Project*, with a design based on a collection of servers called the *servnet*. In Free Haven Project, each server is identified with a *pseudonym* rather than some external identifying information like an IP address. Servers hold pieces of some documents, termed as shares, and exchange these shares in order to promote server anonymity, i.e., no document can be linked back to a server. For further discussion on the various kinds of anonymity provided in the Free Haven design, one can refer to [3] for more on this. The Free Haven design provides a basic framework upon which an efficient reputation system can be built which provides credibility, fairness and accountability.

Accountability is the ability to link an action or a set of actions to the node that is responsible for certain deliverables on time. In any distributed system, providing accountability is a challenging task. Moreover, in a system that is as committed to anonymity as Free Haven, accountability becomes even more challenging. This is because any action by an entity in an anonymous system may be linked to a pseudonymous identity. This pseudonymous identity will most certainly not help establishing any concrete identification of the entity carrying out the actions. The goal of accountability is to maximize a server's effectiveness to the overall system while minimizing its potential (known and plausible) threat [4].

There are two approaches to tackle this problem of minimizing the threat and ensuring accountability. One approach is to limit the risk to an amount roughly equivalent to the benefit from the transaction. This is the basic principle followed by micro-payment schemes. The other approach is to make the risk proportional to the *trust* in the other parties. This forms the base for any reputation system, in which servers are given an incentive to earn reputation by following the protocol, whereas misbehaving leads to a decline in reputation which detriments the server's chances of obtaining services of its peers. Reputations are used extensively in electronic marketplaces to gauge the reliability of online users. The most popular examples of online marketplaces that utilize such reputation systems are eBay and Amazon. A reputation system also consists of scoring algorithms. For example: Any user on eBay can give his counterpart a score of {1, 0, -1} after the transaction. The votes collected by eBay are then used to provide cumulative rating of the user which is made available to all online users. The Free Haven Project provides a strong foundation for reputation system. However, it lacks accountability. Furthermore, the scoring algorithms of servers are also not well-defined.

We propose a reputation system that successfully provides accountability. The proposed scoring algorithm ensures credibility and fairness. Local scores provided by peers are utilized to model cumulative global reputation. The reputation system leverages the use of *authorities* to eliminate feedback broadcasts. The system also introduces a *contract protocol* to be used before the shares trade. A new feedback mechanism is introduced that prevents cheating by the involved parties. The proposed reputation system provides both theoretical foundation and adequate cryptographic safeguards to supplement a Free Haven-like anonymous data storage system.

The remainder of the paper is organized as follows: Section 2 outlines some existing work on reputation systems for peer to peer networks. Section 3 presents our scheme. Section 4 analyzes the proposed scheme. We conclude the paper in Section 5.

## 2 Related Works

Two popular reputation-based systems, CONFIDANT [5] and CORE [6], are designed in such a way that each node is aware of the reputation of every other node in the network. The feedback in CONFIDANT is based on reputation table broadcasts, which is a limitation in terms of performance. CORE on the other hand, has an indirect mechanism, that is, by observing the positive recommendation about other nodes. As mentioned earlier, The Free Haven Project [3] makes an attempt to design a reputation system for their architecture, where some set of ideas have been discussed, but no concrete protocols have been introduced.

Aberer and Despotovik [7] provide a formal analysis of reputation and trust in the context of peer to peer networks. Their approach is based on P-Grid, a decentralized storage method. The probability that an agent will cheat in the future is assessed using the information provided by P-Grid. The proposed system in Section 3 can handle agents having high probability of cheating.

## 3 The Proposed Scheme

We propose reputation algebra for a semi-centralized anonymous data storage system for cloud. The servers trade files with each other to gain reputation. The reputation algebra is divided into six phases: Initial Setup, Registration, Transaction, Scoring and Reputation Algorithms, Revocation of a server, and Change in Authority.

### 3.1 Basic Architecture

The proposed system has a flat-architecture as opposed to a hierarchical one. Nodes and the leaf servers are logically connected to each other. The root servers are authorities that maintain reputations of all leaf servers assigned to them. Both authorities and leaf servers participate in file trading. The authorities change dynamically after periodic timeouts. The reason for periodic timeouts is mentioned later. Consequently, a leaf server is promoted to the role of an authority if its reputation satisfies required criteria. There is a database, which can be accessed by all the authorities, containing information about the node, its global reputation and its authority.

### 3.2 Role of Authority

The root servers in the *servnet* architecture are called the authorities. Authorities are chosen so that they have the highest reputation relative to all leaf servers assigned to it. The *servnet* consists of several interconnected tree-like structures with each sub-tree consisting of an authority and its leaf servers. Since the server having the highest reputation in the sub-tree is chosen as the authority, authorities keep changing. Authorities trade files like leaf servers but also maintain reputations of all the leaf servers connected to them. The reputations are stored in a database and a key is generated on-the-fly by the authority for encryption of the data to be stored in database. This ensures that only the authority can access the database.

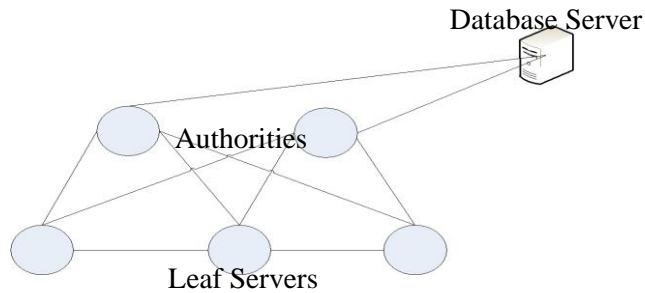

**Figure 1.** *Servnet* Architecture

The notation used in the scheme is given in Table 1.

| Symbol | Meaning |
|---:|---|
| $Auth_X$ | Authority of an entity, say X |
| $Auth_{All}$ | All authorities |
| $Pk_X$ | Public key of X |
| $N_X$ | Nonce of X |
| $sign_x$ | X's signature |
| signAuthC | Chosen Authority's signature |
| $K_X$ | Secret key of X |
| $SS_X$ | Size of the share X |
| $D_X$ | Duration for which X needs to store the file |
| $N_A N_B$ | Transaction ID |
| $\{M\}K$ | M is encrypted by key K |

**Table 1.** Notation used in the scheme

### 3.3 The Scheme
The phases of the scheme are as follows:

**Initial Setup.** The system is deployed with each server assigning score 0 as the initial reputation. The authorities for initial deployment are chosen randomly.

**Registration.** Any server entering to the system needs to register with an authority. The new server is attached as a leaf to an authority. It is assigned 0 as an initial global reputation. The server is identified using a *pseudonym* and its public key. Both these parameters along with an acknowledgement of the authority-server connection are flooded to each authority in the *servnet*. This enables each authority to construct a node-authority table.

The registration phase works as follows:

```
1. A → Auth_C : A, PK_A, {A, PK_A, N_A}_SignA
2. Auth_C → A : {N_A+1}_PkA
3. Auth_C → Auth_All : {A, PK_A, {A, PK_A, N_A}_signA}_signAuthC
```

Here, A indicates a new server and AuthC indicates chosen authority.

**Transaction.** The transaction phase comprises of the Contract Establishment Protocol and the Feedback Protocol.

*The Contract.* Each party sends a message containing the proposed file size and duration for which it is to be stored. If the conditions are agreeable to both, they proceed with the file trade. The Contract Establishment Protocol is shown in Figure 2. Here, *A* and *B* are considered as new servers.

```
1.   A → B : A, SS_A, D_A, B, N_A
2.   B → Auth_B : B, (A, N_B')
3.   Auth_B → Auth_A : Auth_B, {B, A, N_AuthB}_SignAuthB
4.   Auth_A → B : Auth_A, {A, GR_A}_SignAuthA
5a.  B → A: B, Reject, N_A+1
5b.  B → A: B, SS_B, D_B, A, N_B
6.   A → Auth_A : A, (B, N_A')
7.   Auth_A → Auth_B : Auth_A, {A, B, N_AuthA}_SignAuthA
8.   Auth_B → A : Auth_B, {B, GR_B}_SignAuthB
9a.  A → B : A, Reject, N_B+1
9b.  A → B : A, Ack, N_B+1
10.  A → B: {Hash (Messages 1, 5, 9)}_SignA
11.  B → A: {Hash (Messages 1, 5, 9, 10)}_SignB
```

**Figure 2.** Contract Establishment Protocol

*Contract Establishment Protocol Explanation.* *A* sends the request for a share trade with *B* in message 1. The parameters for the trade i.e. Share size and Duration are sent as $SS_X$ and $D_X$ in the message. In steps 2, 3, 4, *B* queries its authority for *A*'s cumulative global reputation. After step 4 is complete, *B* has all the information it needs to make a decision on whether to accept or reject *A*'s trade request.

*B* sends a "Reject" or its own parameters in message 5 (5a or 5b). If a "Reject" is sent, the protocol ends. In message 6, A queries its authority for *B*'s cumulative global reputation. This allows *A* to make a decision to either accept or reject the trade based on *B*'s parameters received in message 5. Again, *A* either sends a "Reject" or an "Ack" based on its decision. The final two messages ensure that *A* and *B* have agreed upon the same parameters that were exchanged in the messages. Once *A* receives message 10, it extracts the hash and compares it with hash of its own set of message 1, 5 and 9 (9a or 9b). If the check fails, protocol is aborted. The same procedure is carried out by *B* with messages 1, 5, 9 and 10 after receiving message 11.

*Feedback.* After every completed transaction, each party sends a mandatory feedback giving a local score of the other transacting party using the Feedback protocol defined below.

*A* gives a local score (LS) to *B* using the Feedback protocol as follows:

$$Feedback_A = A, \{N_A N_B, LS_B, A\}_{SignA}$$

1. A → B: Feedback$_A$
2. A → Auth$_B$: Feedback$_A$
3. B → Auth$_B$: Feedback$_A$

Similarly *B* gives a local score to A

$$Feedback_B = B, \{N_A N_B, LS_A, B\}_{SignB}$$

1. B → A: Feedback$_B$
2. B → Auth$_A$: Feedback$_B$
3. A → Auth$_A$: Feedback$_B$

*Feedback Message Explanation.* Feedback message contains *A*'s identity so that the signature can be verified. The signed part contains $N_A N_B$ i.e. the Transaction ID, local score for *B* given by *A* and the identity of *A*. The identity of *A* is added so that no other server can send a feedback for *B*.

*Feedback Protocol Explanation.* *A* sends its feedback through two paths, which are as follows.

1. A → B → Auth$_B$
2. A → Auth$_B$

Since the feedback passes through *B*, it can extract $LS_B$ and know when *A* gives a false score. Also, *B* cannot modify its score because *A*'s signature on the message cannot be generated. Further reasons are provided in the security analysis of the Feedback Protocol in Section 4.

**Scoring Algorithm.** The scores maintained by an authority is the cumulative global reputation of a server. One of the ingredients to derive the cumulative global reputation is the local score provided in the feedback message at the end of each transaction. The other ingredient is the global reputation of the server that is providing the local score. The local score can be either 1 or -1 based on the result of the transaction. The total cumulative global reputation (GR) of a server is defined as:

$$GR = T * (POS) / (NEG + 1) \quad \text{------------------------- (1)}$$

where,

$$POS = \sum (PS_X * GR_X) \quad \text{----------------------------------- (2)}$$
$$NEG = \sum (NS_X * GR_X)$$

The global reputations of *A* and *B* are derived in the following manner after the feedback is received:

$$GR_A = (N_A + 1) * (POS_A) / (NEG_A + 1*GR_B + 1)$$

$$GR_B = (N_B + 1) * (POS_B + 1*GR_A) / (NEG_B + 1)$$

As mentioned earlier, the three major properties that need to be satisfied by the reputation system are accountability, credibility and fairness. Below we show that the proposed scoring algorithms ensure accountability, credibility, fairness.

*Accountability.* Accountability is inherently provided with the reputation system. Any peer that misbehaves will be caught and a negative point will be sent to the authority thereby reducing its reputation. Reduced reputation decreases its chances of making further transactions. Therefore, a misbehaving server cannot sustain itself in the system.

*Credibility.* The credibility of a server is dependent on its global reputation. Higher global reputation implies higher credibility. This ensures that a new server cannot degrade the reputation of an existing server by giving negative scores.

In (2) and (3) positive and negative local scores by a server *X* are multiplied with global reputation of *X*, i.e., the weight of *X*'s score depends on the global reputation of *X*. For example, a negative score by a low-reputation server does not cost the other party much; however, a positive score by a high-reputation server carries a lot more weight.

*Fairness.* Consider a server which is performing consistently and has lower number of transactions as compared to a relatively inconsistent performer with higher number of transactions. Fairness ensures that after a certain number of transactions, the consistent performer has a higher global reputation than the inconsistent server. Let there be two peers, peer 1 and peer 2. Let $T_1$ and $T_2$ be the number of transactions by peer 1 and peer 2, respectively in a particular time frame. Also let peer1 receive a negative score after every $m_1$ transactions and $m_2$ be the number of transactions after which peer 2 receives a negative score. The condition for which peer 1's reputation is greater than peer 2's reputation is as follows:

We want $GR_1 > GR_2$

$$T_1 * T_1(m_1-1) / m_1*(T_1/m_1+1) > T_2 * T_2(m_2-1) / m_2*(T_2/m_2+1)$$
$$T_1 * T_1(m_1-1) / (T_1+m_1) > T_2 * T_2(m_2-1) / (T_2+m_2)$$
Let $k = T_2 * T_2(m_2-1) / (T_2+m_2)$
$$T_1^2(m_1-1) > kT_1 + m_1 k$$
$$T_1^2(m_1-1) - kT_1 - m_1 k > 0$$

Solving for $T_1$,
$$T_1 > (T_2(m_2-1)/T_2+m_2 \pm (k^2+4(m_1-1)m_1k)^{1/2})/2(m_1-1)$$

If the above condition is satisfied, the reputation of peer 1 is greater than that of peer 2, else the reputation of peer 2 is greater than that of peer 1.

We use (1) to calculate the cumulative global reputation of each peer using the parameters mentioned in the Table 2.

| Parameters | Peer 1 | Peer 2 |
|---|---|---|
| No. of transactions | $T_1$ | $T_2$ |
| No. of transactions after which a negative score is Received | $m_1$ | $m_2$ |
| Current Negative Score | $T_1/m_1$ | $T_2/m_2$ |
| Current Positive Score | $T_1(m_1-1)/m_1$ | $T_2(m_2-1)/m_2$ |

**Table 2.** Formulas for peer 1 and peer 2

**Revocation of a server.** The server before leaving the *servnet* transfers the data that is part of a contract to its respective owners and the remaining data is transferred to its authority. The reputation for the server is set to 0 and all other authorities are notified.

**Change of Authorities.** After a certain timeout, a mandatory mechanism to change the authority is invoked. When the timeout occurs, there will be a change in authority under the following circumstances:

a) The global reputation of the contender is higher than the global reputation of the current authority.
b) Global reputation of the contender is higher than some factor of the global reputation of the current authority. For example: Let that factor be 1/2. Therefore, if

(Global reputation of contender)>1/2*(Global reputation of current authority)

the contender will be appointed as the new authority. The factor may vary according to owner policies.

If the conditions a) and b) are not satisfied then the current authority continues to be the authority and it will remain as authority till any of the above two conditions are satisfied.

If an authority change does take place, the old authority loses access whereas the new authority gains access to the database server. Also, the database is updated to reflect the authority change. Each node attached to the old authority is notified of the authority change. The following Authority Change Protocol is used to generate a new key to access the database server. This is carried out in such a way that the old authority is denied access to the database.

1. $Auth_{Old} \rightarrow$ DB Server: $\{B, A, N_A\}_{KDBA}$
2. DB Server$\rightarrow Auth_{New}$: DBServer,$\{\{K_{New}, B, N_{DB}\}_{SignDB}\}K_B,\{Auth_{New}, Auth_{Old}, N_{DB}\}_{SignDB}$
3. $Auth_{New} \rightarrow$ DB Server: $\{N_{DB} + 1\}_{KNew}$

Each leaf server attached to the old authority is notified about the change in authority by sending the following message.

$$DBServer, (Auth_{New}, Auth_{Old}, N_{DB})_{SignDB}$$

The terms in this message can be found in message 2 of the Authority Change Protocol. Also, this message cannot be generated by any authority on its own due to DB server's signature. The Authority Change Protocol is run when an authority leaves the *servnet*.

## 4   Analysis of the Scheme
This section contains security and performance analysis of the proposed scheme.

### 4.1   Contract Establishment Protocol
*Impersonation.* A malicious party may attempt to impersonate another peer to establish a contract using the proposed protocol. The impersonating party may deliberately fail to fulfill the contract conditions. This results in a negative feedback for the impersonated peer. In the proposed protocol (Figure 2), all the fields in the message $A$, $TS_A$, $SS_A$, $D_A$, $B$, $N_A$ can be generated by the attacker but the signatures in messages 10 and 11 cannot be generated. Therefore, the impersonation attack fails.

*Man-in-the-middle attacks* (MITM). An adversary between $A$ and $B$ can modify $SS_A$ and $D_A$ in the first message or $SS_B$ and $D_B$ in the second message in such a way that the peer on the other end of the transaction will not agree with them. In the proposed protocol, even if such a change is made by generating custom values of share (SS), the adversary cannot generate the correct hashes sent in messages 10 and 11. Without the correct hashes, the hash checking on both ends fail. Hence MITM is prevented.

*Replay attack.* Replay attacks require the fraudulent transmission of valid transmission data. In our protocol (Figure 2), the presence of nonce in messages 1 through 9 prevents them from being replayed without being detected. Messages 10 and 11 are just signed hashes of messages (1, 5 and 9) and (1, 5, 9 and 10) respectively. Replaying messages 10 and 11 will not be fruitful because the hash checking at each end will fail.

### 4.2   The Feedback Protocol
*Impersonation.* A peer may impersonate another peer to give a false score to its contracting party. In the proposed protocol, the impersonating party cannot generate the signature over the term $(N_A N_B, LS_B, A)_{signA}$ even though the individual terms can be calculated. Hence impersonation attack fails.

*Party giving the score cannot cheat.* As the protocol suggests, $A$ gives a feedback for $B$ throughout two distinct paths. $A$ sends the feedback message to $B$ who then forwards it to his authority. A also sends the feedback message directly to $B$'s authority. In the first path, $B$ can check whether $A$ has given a false score. Also, if $A$ gives a positive score in the feedback message sent to $B$ and a negative score in the feedback message directly sent to $B$'s authority, the discrepancy can be detected. Since $A$ signs both feedback messages, the authority can know that $A$ is cheating.

*Party receiving the score cannot cheat.* Suppose $B$ receives a negative score from $A$ and decides to drop the feedback message, i.e., not forward the feedback message to its authority. Since $A$ sends a feedback message to $B$'s authority directly as well, this problem can be resolved. Therefore, $B$ cannot cheat.

### 4.3 Authority Change Protocol

*Impersonation.* A normal peer cannot impersonate an authority during the run of the "Change in Authority" protocol. This is because the first message (B, A, $N_A$) is encrypted using $K_{DBA}$ which is known only to the old authority and the database server. A normal peer or an authority cannot generate the signature. Furthermore, the last message $\{N_{DB} + 1\}_{KNew}$ ensures that the new authority knows the correct key.

## 5 Conclusion

We proposed a scheme for reputation system built atop the Free Haven framework for anonymous cloud data storage system. Our scheme provides dynamic operation such as adding servers into the system, removing servers from the system, establishing trade contracts between two parties, feedback mechanism for exchanging local trust scores and changing authorities after mandatory timeout. The proposed scoring algorithm provides accountability, credibility and fairness, which are the most important attributes of any reputation system. The analysis of sub-protocols shows that the proposed scheme is secure against known and plausible threats.